\documentclass[aps,prl,floatfix,twocolumn,superscriptaddress,showpacs,epsf]{revtex4-1}

\usepackage[USenglish]{babel}
\usepackage{amsmath,amssymb,amsfonts}
\usepackage{epstopdf}
\usepackage{epsfig} 
\usepackage{graphicx}
\usepackage{subfigure}
\usepackage{import}
\usepackage{color}
\usepackage{url}

\allowdisplaybreaks

\usepackage[colorlinks=true,linkcolor=blue,citecolor=red]{hyperref}

\newcommand{\ca}{c^{\phantom{\dagger}}}
\newcommand{\cc}{c^\dagger}

\newcommand{\hcc}{\hat{c}^\dagger}
\newcommand{\hca}{\hat{c}^{\phantom{\dagger}}}
\newcommand{\be}{\begin{equation}}
\newcommand{\ee}{\end{equation}}
\newcommand{\bea}{\begin{eqnarray}}
\newcommand{\eea}{\end{eqnarray}}
\newcommand{\ba}{\begin{eqnarray*}}
\newcommand{\ea}{\end{eqnarray*}}

\newcommand{\bk}{\mathbf{k}}

\newcommand{\kpar}{\mathbf{k}_{\|}}

\newcommand{\eqn}[1]{(\ref{#1})}

\def\ie{\mbox{\it i.e.\ }}

\begin{document}

\title{Field-driven Mott gap collapse and resistive switch in correlated insulators}
\author{G.~Mazza}
\affiliation{Scuola Internazionale Superiore di Studi Avanzati (SISSA),  
Via Bonomea 265, 34136 Trieste, Italy}
\affiliation{Centre de Physique Th\'eorique, \'Ecole Polytechnique, CNRS, Universit\'e Paris-Saclay, 91128 Palaiseau, France}
\affiliation{Coll\`ege de France, 11 place Marcelin Berthelot, 75005 Paris, France}
\author{A.~Amaricci}
\affiliation{Scuola Internazionale Superiore di Studi Avanzati (SISSA),  
and Democritos National Simulation Center, 
Consiglio Nazionale delle Ricerche, Istituto Officina dei Materiali (CNR-IOM), 
Via Bonomea 265, 34136 Trieste, Italy}
\author{M.~Capone}
\affiliation{Scuola Internazionale Superiore di Studi Avanzati (SISSA),  
Via Bonomea 265, 34136 Trieste, Italy}
\author{M.~Fabrizio}
\affiliation{Scuola Internazionale Superiore di Studi Avanzati (SISSA),  
Via Bonomea 265, 34136 Trieste, Italy}


\begin{abstract}
  Mott insulators are ``unsuccessful metals"  in which Coulomb repulsion prevents charge
  conduction despite a metal-like concentration of conduction electrons. 
  The possibility to unlock the frozen carriers with 
  an electric field  offers tantalizing prospects of realizing 
  new Mott-based microelectronic devices. Here we unveil how such
  unlocking happens in a simple model that shows coexistence of a
  stable Mott insulator and a metastable metal. 
  Considering a slab subject to a linear potential drop we find, by
  means of Dynamical Mean-Field Theory
  that the electric breakdown of the Mott insulator occurs via a first-order
  insulator-to-metal transition characterized by an abrupt 
  gap-collapse in sharp contrast to the standard Zener breakdown. 
  The switch-on of conduction is due  to the field-driven stabilization of the metastable metallic phase.
  Outside the region of insulator-metal coexistence, the electric breakdown 
  occurs through a more conventional quantum tunneling across the Hubbard bands tilted by the field. 
  Our findings rationalize recent experimental  observations and may offer a guideline for future technological
  research. 
\end{abstract}

\maketitle

\paragraph*{Introduction.} 
The conventional description of the electric breakdown, i.e. the field driven formation of a
conductive state in an otherwise insulating system, is based on the
well-known Landau-Zener mechanism of quantum tunnelling across the
insulating gap~\cite{zener_original,Landau1932}. 
However, the existence of a ``rigid'' band gap to be overcome by the
field  sets a lower bound of the threshold field and 
limits the density of excited carriers promoted
across the gap.
This ultimately leads to large density fluctuations at the nanoscale,
one of the bottlenecks in device miniaturization~\cite{Newns}. 
These limitations are intrinsic in band insulators,
whose gap is fixed as long as chemical composition and lattice structure do not change. 

Because of the collective nature of their gap, Mott insulators have
recently emerged as ideal candidates to overcome the above issues,
providing a potential alternative to semiconductor-based
microelectronics \cite{Newns,Takagi}.  
The possibility of driving a gap closure in a Mott
insulator by means of an electric field could enable to access
a much larger carrier density than in semiconductors, potentially 
overcoming most limitations of conventional devices.
\begin{figure*}
  \centering\includegraphics[width=1\linewidth]{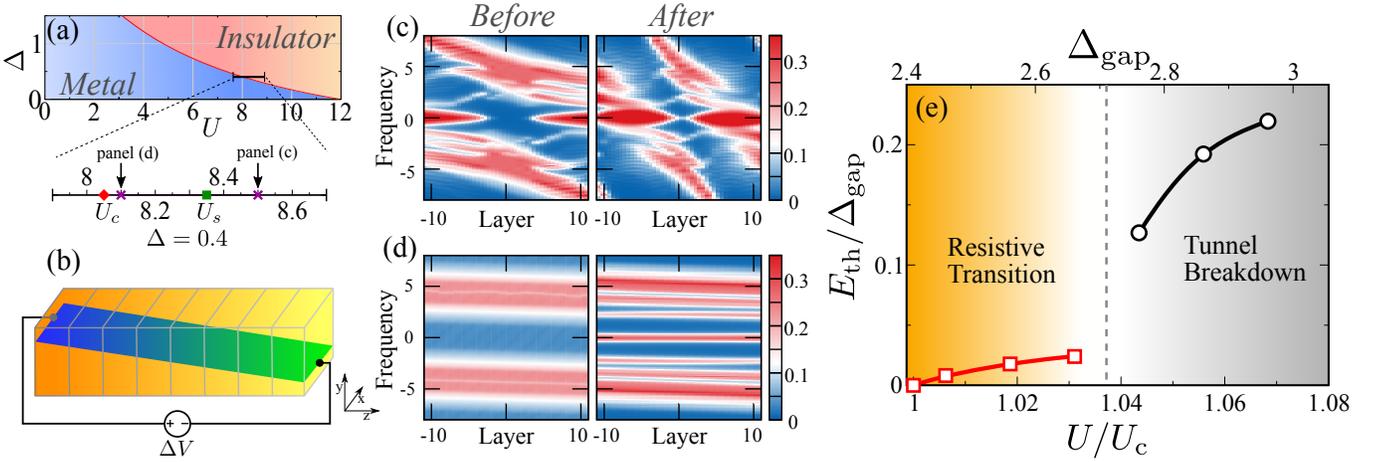}
  \caption
      {
        (Color online) 
        (a) {\em Zero-field phase diagram.} 
        Schematic representation of the equilibrium phase diagram.
        The blow-up at $\Delta=0.4$ highlights the range of the
        interaction relevant for this study. 
        $U_c$ marks the Mott transition critical value (red
        diamond). $U_s$ marks the spinodal point (green square).
        The arrows indicate the interaction strengths used in panels
        (c) and (d) (purple crosses). 
        (b) \emph{Sample geometry}. The sample is a layered slab
        subject to a linear voltage drop $\Delta V$, corresponding to a
        uniform and static electric field along the slab direction.      
        (c-d) \emph{Electric-field induced insulator-to-metal transition}.
        Layer-resolved local spectral densities before (left) and
        after (right) the field driven
        insulator-to-metal transition for two values of the
        interaction strength: $U\!=\!8.5$ (c) and $U\!=\!8.1$ (d), 
        respectively outside and inside the coexistence region. The
        intensities of the applied fields are 
        $E\!=\! 0.2$ (left) and $E\!=\!0.6$ (right) for panels (c) and 
        $E\!=\!0.01875$ (left) and $E\!=\!0.025$ (right) for panels (d). 
        (e) \emph{Electric-field Vs. interaction phase diagram}. 
        Threshold field $E_\text{th}$ in units of the zero-field
        insulating gap $\Delta_\text{gap}$ versus $U/U_c$. 
        We also show $\Delta_\text{gap}$ for each value of $U/U_c$
        (upper x-axis).        
      }
      \label{fig1}
\end{figure*}

Experimental evidences for such an appealing field-driven {\it resistive
switch} have been recently found in several Mott insulators
~\cite{cario_NatComm,Cario_AdvMat2013,Kim-VO2-2010,Nakamura-Ca2RuO4-2013,Natelson-Fe3O4-2009,Natelson-Fe3O4-2012,Guenon-V2O3-2013,Janod-2015} 
and Mott-based devices~\cite{iwasa_VO2,Iwasa_ApplPhysLett}.
Remarkably, these experiments ubiquitously report a whole novel
scenario for the electric breakdown that cannot be reconciled with 
the standard Landau-Zener description. For instance, the breakdown occurs through 
the abrupt formation of a conductive channel at 
anomalously small threshold fields~\cite{cario_NatComm}, as opposed to
the smooth activation at fields of the order of the gap as predicted by across-gap tunnelling. 
In addition, VO$_2$ electric double layer transistors formed at a
solid/electrolyte interface show massive conducting channels that
suddenly open up above a threshold gate-voltage, with an extension
much beyond the fundamental electrostatic screening
length~\cite{iwasa_VO2,Iwasa_ApplPhysLett}. 

While several mechanisms of resistive switch beyond the tunnelling paradigm are known, including thermally driven electric breakdown
~\cite{joule_heating_breakdownVO2,Brockman-V2O3-2014}, valence-change
driven insulator-to-metal transitions~\cite{NiO_FET,CuO_FET} and
resistive transitions generated by the non-linear propagation of ionic defects~\cite{shock_waves_RS}, we 
believe that the above experiments still suggest another possible
scenario of a genuine insulator-to-metal Mott transition triggered by
the external field, as proposed in
Refs.~\cite{cario_NatComm,Cario_AdvMat2013} and modelled
phenomenologically through a resistor network mimicking the
competition between a stable insulating phase and a metastable
metallic one. 
Nonetheless a true microscopic description in model Mott insulators is
still lacking.
Indeed, theoretical studies within the single-band Hubbard model, the
paradigm of strongly correlated systems, have so far highlighted a
breakdown that is essentially due to the tunnelling across the Mott
gap, as if the latter were as rigid as the band gap in
semiconductors~\cite{Oka2003,SOkamoto_PRL,eckstein_dielectric_breakdown,JLi_resistive_switch,gmazza_PRB}.
Even if in some cases the tunnelling breakdown can be anticipated by
the formation at large fields of in-gap states due to the
Wannier-Stark
effect~\cite{Joura_steady_state2008,Tsuji2008PRB,LeePark_WannierStark},
still the agreement with the above experiments remains poor. 

Motivated by the strong evidence that the first-order character of the
Mott transition plays a major role in the aforementioned
experiments~\cite{cario_NatComm,iwasa_VO2,Cario_AdvMat2013}, in this
Letter we explore the route to the electric breakdown that opens
whenever a stable Mott insulator coexists with a metastable metal
which is not connected to the insulating solution.

We realize this situation including extra degrees of freedom to the Hubbard model~\cite{Inoue,Matteo2015},
and we choose an orbital degree of freedom which is ubiquituously
relevant in actual Mott insulating materials and increases the
coexistence region with respect to the single-orbital model. 
In particular we consider the simplest modelling of a Mott insulator with a $d$-$d$ gap~\cite{ZS&A1985},  
which we study in a slab geometry and in the presence of a constant
electric field, \ie an open circuit setup mimicking a
FET~\cite{iwasa_VO2} with a gate voltage but without source-drain
bias.


We show that, within the insulator-metal coexistence region, an
electric field can drive a discontinuous transition from the insulator to a
gap-collapsed metal at threshold fields much smaller than
those expected in a Zener breakdown. 
%

\paragraph*{Model.}
We consider a half-filled two-orbital Hubbard model in a three dimensional lattice and in the presence of a crystal-field $\Delta>0$ that lifts the orbital degeneracy. The generic Hamiltonian reads  
\begin{equation}
  \begin{split}
    \mathcal{H} =& \sum_{\bk \sigma} \sum_{\alpha,\beta =1}^2
    t_{\bk}^{\alpha \beta}\, \cc_{\bk \alpha \sigma} \,\ca_{\bk \beta \sigma} 
    - \frac{\Delta}{2}\, \sum_{i} \,\big( n_{i,1} - n_{i,2} \big) \\
    \phantom{=}& + \frac{U}{2} \,\sum_i \,\big( n_i - 2 \big)^2,
  \end{split}
\label{MainModel}
\end{equation}
where $n_{i,\alpha}\! =\! \sum_{\sigma}\! \cc_{i \alpha \sigma} \ca_{i \alpha \sigma}$ and
$n_i\! =\! \sum_{\alpha}\! n_{i,\alpha}$ are density operators at site
$i$. At $U=0$ and for not too large $\Delta$ the model
\eqn{MainModel} describes a two band metal. 
At large $U$ the ground state is a non-magnetic Mott insulator, 
with one filled and one empty orbital. Such insulating state is stable
against spin and/or orbital ordering due to the finite value of
$\Delta$. 
This is in sharp contrast with the single-band Mott insulator
description, in which the extensive spin entropy inevitably favors
the onset of a magnetic order. 
Although the presence of a symmetry broken phase at weak or
intermediate coupling in model \eqn{MainModel} can not be
excluded, its very existence and its properties critically depends on
the model details. 
Therefore, to maintain the discussion as general as
possible, in what follows we shall not account for possible broken
symmetry phases~\cite{JLi_AFM_resistive_switch}.

In order not to weigh the Hamiltonian with too many parameters, 
we choose for our analysis   
$t_{\bk}^{11}\! =\! t_{\bk}^{22}\! =\! 
-2t\,( \cos k_x + \cos k_y + \cos k_z )$ the intra-orbital dispersion
on a three-dimensional cubic lattice and $t_{\bk}^{12}\!=\! t_{\bk}^{21}\! =\! v \,(\cos k_x -\cos k_y)
\cos k_z$ a non-local hybridization that leaves the local
single-particle density matrix diagonal in the orbital index even though the occupation of each orbital is not
a conserved quantity. 
We observe that, because of our very specific choice, the model possesses an orbital U(1) symmetry that can be broken
by a Stoner instability at weak coupling \cite{exciton_insulator}.
However, as we previously mentioned, we shall restrict our analysis only to the symmetry invariant subspace.  
The energy unit is such that the intra-orbital hopping is $t\!=\!0.5$ and we take $v\!=\!0.25$. We solve the model by means of dynamical mean-field theory (DMFT)~\cite{Review_DMFT_96,potthoff_rDMFT} using an exact
diagonalization solver (see Supplementary Materials).

Under the above assumptions, the model \eqn{MainModel} undergoes a first-order Mott transition at a critical value of
$U\!=\!U_c$ monotonically decreasing with increasing $\Delta$ (see Fig.\ref{fig1}(a)).
For the sake of definiteness, in the following we fix the
crystal-field splitting to $\Delta = 0.4$, for which
$U_c\!\simeq\!8.05$. 
For $U_c < U < U_s\!\simeq\!8.3$ the insulating solution is stable,
but the metal continues to exist as a metastable solution up to the spinodal point $U_s$.
The state variable that better characterizes the transition is the
orbital polarization $m\!=\!n_1\!-\!n_2$, \ie the population imbalance
between lower and upper orbital. 
At $U\!=\!0$, the model describes a partially polarized metal
($m\!<\!2$). 
A finite interaction $U$ reduces the effective bandwidth and induces a repulsion between occupied and unoccupied states, leading to an effective enhancement of the crystal field $\Delta_\mathrm{eff}\!>\!\Delta$ that increases the
orbital polarization $m$.  
At the first-order transition the metal turns abruptly into an almost  fully polarized
insulator ($m \approx 2$)~\cite{werner_millis_2bands,poterayev_2bands}, 
with a finite gap separating the occupied lowest band from the empty upper one~\cite{supplementary}. 
A sort of Mott insulator ``disguised'' as a conventional band insulator.

In order to study the effect of an applied electric field we consider
a layered slab of our idealized material subject to a static and
uniform electric field $E=\Delta V/N$ (Coulomb gauge) directed along
the slab direction (see Fig.\ref{fig1}(b)), namely an open circuit configuration with $\Delta V$ playing the role of a gate voltage.

\paragraph*{Results.} 
Starting from the equilibrium insulator at $U\!>\!U_c$ we 
increase the electric field $E$ and monitor the ground state evolution until, above a threshold value $E_\text{th}$, a {\sl conducting} state with a
finite density of states at the Fermi level is established throughout the slab. 
We find that such field-induced conducting state has strikingly different
properties depending whether the insulator at zero field is within the coexistence region, $U_c<U\lesssim U_s$, 
or outside, $U>U_s$.  
This is highlighted by panels (c)-(d) of Fig.\ref{fig1} showing the electric-field dependence
of the layer resolved spectral density outside, panel (c), or inside, panel (d), the coexistence region. 
In both cases we show the spectra immediately before and after the formation of the conducting state.

Outside coexisistence, i.e.  $U\!>\!U_s$ (panel c), we observe a marked tilting of lower (LHB) and upper (UHB) Hubbard bands, 
just as we would expect in a band insulator. When the field is large enough, the LHB crosses the Fermi energy at the left side of the slab, 
which is thus doped with holes, while the UHB does the same at the right side, which is instead doped with electrons. 
As a result, two facing and oppositely charged surface-layers with finite spectral weight at the Fermi level appear~\cite{supplementary}, 
whose thickness grows with the field until, above the threshold value $E_\text{th}$, they  touch in the center of the slab. 
Before touching, the two surface-layers are tunnel-coupled through an insulating barrier,  the central part of the slab~\cite{supplementary}, and only above $E_\text{th}$ 
a sizeable density of states at the Fermi level is established throughout all the sample. 
We note that this behaviour is essentially that originally uncovered by Zener~\cite{zener_original} in the Coulomb gauge, though here in an open circuit configuration, 
as already discussed in the case of Mott insulators~\cite{Oka2003,pashkin_VO2}.

A completely different behaviour arises instead when the system is perturbed within the coexistence region between metal and insulator ($U_c\!<\!U\!\lesssim\!U_s$). 
In this case a sharp insulator-to-metal transition instead takes place at the threshold field $E_\text{th}$. 
This is illustrated by the abrupt change of the
layer-resolved spectral density across the resistive transition
reported in Fig.~\ref{fig1}(d).  As the threshold field is 
crossed, the gap abruptly collapses and a sizeable spectral weight at
the Fermi level appears. The absence of any precursor on the insulating side 
indicates that the system displays a true resistive switch in which the ground state 
sharply changes from an insulator to a metal.
As opposed to the previous case, the field-induced metal is homogeneous
with a sizable spectral weight quite uniform throughout the whole slab~\cite{supplementary}.

In panel (e) we plot the threshold field
$E_\mathrm{th}$ in units of the zero-field insulating gap
$\Delta_\text{gap}$ as a function of the distance from the first-order Mott transition. 
A clear break is observed in the curve around $U \simeq 1.04 U_c\simeq U_s$
separating the two distinct regimes. The threshold field
is small and slowly increasing within the coexistence region, 
while it gets larger and more steeply increasing outside. 
Remarkably, the strong variation of $E_\text{th}$ is in sharp contrast with the weak variation of 
the gap.

\begin{figure}
  \begin{center}
    \includegraphics[width=1\linewidth]{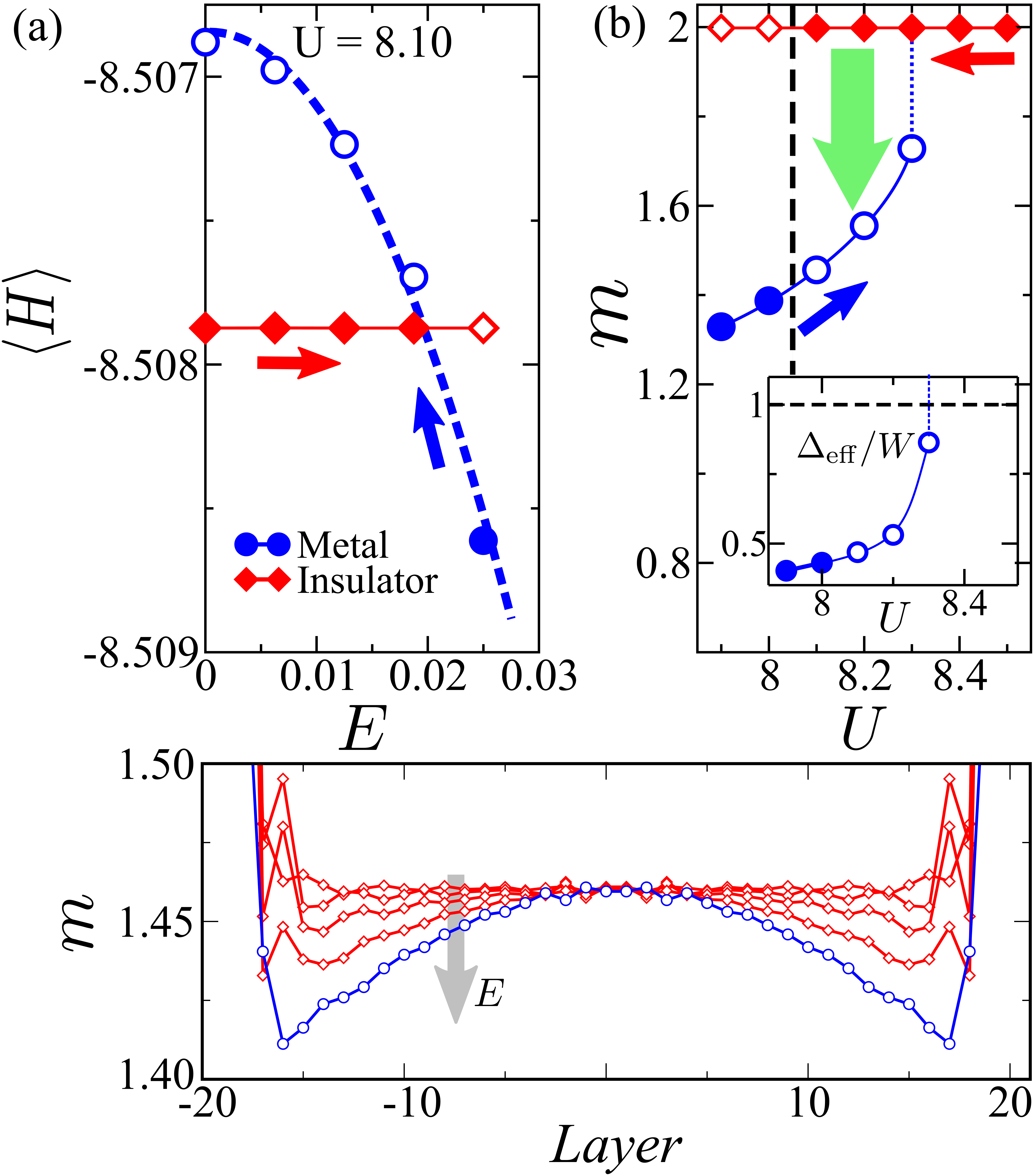}
  \end{center}
  \caption{(Color online) 
    \emph{Metal-insulator coexistence}. 
    (a) Internal energy $\langle H\rangle$ of the metallic (blue
    circles) and the insulating (red diamonds) 
    solutions, as a function of the electric field $E\!=\!\Delta V/N$. 
    The filled/open symbols mark the stable/metastable character
    of the solution for each value of $E$. 
    Dashed blue line is a quadratic fit to the energy of the metallic solution (see text). Red full line is a guide to the eye.
    (b) Hysteresis loop for the orbital polarization $m\!=\!n_1\!-\!n_2$ across the zero-bias 
    Mott transition.  
    The red and blue arrows define the directions of continuous evolution for the insulating and metallic phases 
    respectively. The big green arrow indicates the electric-field induced switch between the stable insulator
    and the metastable metal. Lines are guides to the eye.
    \emph{Inset}: Effective crystal field
    $\Delta_\text{eff}$ in units of the bandwidth $W$ as a
    function of the correlation strength $U$. The metallic state is destabilized for $\Delta_\mathrm{eff} > W$.
    (c) Polarization profile for the metastable metal for increasing electric field strength (grey arrow).
    Red diamonds/blue circles refer to a system with an insulating/metallic ground state.
  }
  \label{fig2}
\end{figure}

It is evident that the low-field resistive transition not extending beyond the spinodal point $U_s$ cannot be accidental. 
Indeed, as neatly shown in Fig.~\ref{fig2}(a), the field-induced metallic phase is adiabatically connected with the metastable metal at zero field. 
At $E\!=\!0$ the insulating solution has lower internal energy and is
separated from the metastable metal by a small energy difference.
However, the metallic solution has a high electrical polarizability and it
gains more energy from the field $E$ than the incompressible 
Mott insulator, whose energy remains essentially constant.   
As a consequence, the two energy curves eventually cross at the threshold field
$E_\text{th}$. The first-order character of such field-driven insulator-metal transition is further highlighted  
by the hysteresis loop of the equilibrium orbital polarization $m$ 
across the Mott transition, plotted in Fig.~\ref{fig2}(b). 
The electric field thus acts as a switch between two phases with no adiabatic connection across 
the first-order metal-insulator transition~\cite{camjayi_GaTaSe}.

Despite the field $E$ has the dramatic role of driving the first order insulator-metal transition, from the viewpoint of the metastable metal 
it is just a weak perturbation whose effects are accounted by linear response theory. 
Indeed, the energy of the metastable metal is well fitted with a simple parabola, see Fig.~\ref{fig2}(a), as predicted by linear response theory, 
$\langle H\rangle(E) = \langle H\rangle(0) - \chi\,E^2/2$, where $\chi$ is the polarisability whose estimate is $\chi\sim 5.46$. 
This is important for what concerns the relevance of our results once the circuit were closed or a weak probing bias applied e.g. perpendicularly to the slab axis. Indeed, in the linear response regime we do not expect any prominent effect caused by a finite current flow~\cite{gmazza_PRB} that could alter completely the physics with respect to the open circuit case. 

In Fig.\ref{fig2}(c) we show that, as long as the system is within the coexistence region ($U<U_s$), the energy gain is accompanied by a significant
reduction of the value of $m$ in the metastable metal, still in the linear response regime apart from finite size effects at the boundaries.
This is remarkably since it appears without a strong charge redistribution across the sample~\cite{supplementary}.
In other words, the net effect of the electric field is essentially to decrease the orbital polarization of the metallic state, 
effectively moving in the phase diagram of Fig.\ref{fig1}(e) as if $U$ or  $\Delta$ were reduced. 
In such a situation the threshold field depends only from the metal polarizability and we expect that it is remains constant as a function of the slab length until the potential at the boundaries becomes so large that non-linear effects appear so that coupling with iso-potential sources might become crucial to observe the same physics;  an event we do not explore here. 

Outside the coexistence region ($U>U_s$), where a metastable metal no longer exists, the field modifies the insulating state through a strong charge redistribution, 
ultimately leading to the formation of the highly inhomogeneous conductive state in Fig.~\ref{fig1}(c). 
The closure of the circuit in this case will induce more dramatic effects, even though we still expect the overall picture 
of a Landau-Zener tunnel breakdown to hold, possibly accompanied by other emergent effects~\cite{Tsuji2008PRB,LeePark_WannierStark}. 

\paragraph*{Conclusions.}
The above results unveil a so far unexplored pathway to the metallisation of a correlated insulator, where the electric field drives a first-order transition from the insulator to a gap-collapsed metal phase, pre-existing as metastable state at zero field. Although the condition of being inside the insulator-metal coexistence region might be considered a rare circumstance in reality, we note that that observed Mott transitions are often 
characterised by quite wide hysteresis loops, see e.g. \cite{Limelette89} in the popular case of V$_2$O$_3$.  

Moreover, those results help clarifying why a similar phenomenon has not yet been observed, at least within DMFT, 
in the single-band Hubbard model, despite the latter also shows metal-insulator coexistence. 
In this model in the whole coexistence region the metal differs from the insulator only for the presence of a
narrow quasiparticle peak at the Fermi level that hosts only
a tiny percentage of carriers. 
For the same reason, the region of parameters where a stable insulator coexists with a metastable metal is extremely narrow,
actually vanishing at zero temperature, so that one ends up   
observing always the same behaviour in a field as that of our model beyond the spinodal line~\cite{supplementary}.

Notwithstanding the obvious importance of the size of the coexistence domain
on the insulating side of the Mott transition, we believe that the key issue to observe 
a genuine resistive switch is the competition of two neatly
distinguishable phases, characterized by sharply different values of
an extensive observable, in our case the orbital polarization $m$,
which can be tuned by the external field.
In turn, this property might also be the rationale of 
the abrupt gap collapse at the resistive transition, which reflects
the very nature of the gap~\cite{Matteo2015}. In our Mott
insulator model the gap separates two bands of different orbital character,
\ie it refers to the cost of redistributing electrons among the
valence orbitals without changing valence, as opposed to the gap
between lower and upper Hubbard bands in an idealized Mott insulator,
which refers instead to the cost of changing the total charge within
the valence shell. 

A similar mechanism could be active in several known Mott insulators
where the ``polarization'' of some degree of
freedom starkly distinguishes the metal from the insulator.
The physical nature of such observable can be
material dependent -- for instance in the case of VO$_2$~\cite{Goodenough1971,Eyert2002,VO2-Georges} or V$_2$O$_3$~\cite{Georges&Andersen,Daniel} 
it is supposed to be the relative occupancy of two different $d$-orbitals, much alike our simple case-study --  yet the correlation-driven Mott
transition would be associated to a sharp change of its value. In such context, the application of an external field would change this ``polarization'' favouring the metallic state and ultimately
driving the resistive transition. 


\paragraph*{Acknowledgements.}
We thank A. Georges, E. Janod, A. J. Millis, M.J. Rozenberg,  for insightful discussions. 
A.A. and M.C. are financed by the European Union under FP7 ERC Starting
Grant No. 240524 ``SUPERBAD''.  Part of this work was supported
by European Union under the FP7 and H2020 Framework Programmes, Grant
No. 280555 ``GO FAST'' and ERC Advanced Grant No. 692670 ``FIRSTORM", respectively. 
G.M. acknowledges support of the European Research Council (ERC-319286 ``QMAC'').

\clearpage
\onecolumngrid

\setcounter{figure}{0}

\begin{center}
  {\bf Supplementary Informations: \\Field-driven Mott gap collapse and resistive switch in correlated insulators}\\
  G. Mazza, A. Amaricci, M. Capone and M. Fabrizio
\end{center}

\onecolumngrid

\section{Model and Method}

We consider the following two-bands model Hamiltonian:
\begin{equation}
  \begin{split}
    \mathcal{H} =& \sum_{\bk \sigma} \sum_{\alpha,\beta =1}^2
    t_{\bk}^{\alpha \beta}\, \cc_{\bk \alpha \sigma} \,\ca_{\bk \beta \sigma} 
    - \frac{\Delta}{2}\, \sum_{i} \,\big( n_{i,1} - n_{i,2} \big) + \frac{U}{2} \,\sum_i \,\big( n_i - 2 \big)^2,
  \end{split}
\label{MainModel}
\end{equation}
where $n_{i,\alpha}\! =\! \sum_{\sigma}\! \cc_{i \alpha \sigma} \ca_{i \alpha \sigma}$ and
$n_i\! =\! \sum_{\alpha}\! n_{i,\alpha}$, while  
$t_{\bk}^{11}\! =\! t_{\bk}^{22}\! =\! 
-2t\,( \cos k_x\! +\! \cos k_y\! +\! \cos k_z )$ is the intra-band dispersion
on a three-dimensional cubic lattice. We also add a non-local
hybridization $t_{\bk}^{12}\!=\! t_{\bk}^{21}\! =\! v \,(\cos k_x -\cos k_y)
\cos k_z$ which allows for inter-orbital charge fluctuations but
leaves the local single-particle density matrix  diagonal in the
orbital index. We set the energy unit such that
the hopping $t\!=\!0.5$, the hybridization $v\!=\!0.25$.

In addition we consider a constant electric field 
directed along the $z$-axis:  $\vec{E}=E\vec{z}$. 
Working in the Coulomb gauge we express the
electric field in terms of a linearly varying potential:
\begin{equation}
  V(z) = V_0 - E z
\end{equation}
and we assume that the field is imposed on a system with finite
extension along the $z$-direction, mimicking a sample between two
external leads kept at finite voltage difference $\Delta V$. To this
extent we introduce a three-dimensional layered structure (Fig.~\ref{fig1}(b) main text)
and fix the reference potential value $V_0$ imposing
a symmetric voltage drop $\Delta V/2$ respect to its center:
\begin{equation}
  V(z) = -\frac{\Delta V}{2} + \Delta V \frac{z-1}{N-1}.
  \label{eq:scalar}
\end{equation}

The Hamiltonian for the slab structure is obtained performing a
discrete Fourier transform along the $z-$direction of the fermionic
operators defined in momentum space 
\begin{equation}
  \cc_{\bk_\| z \alpha \sigma} = \sqrt{\frac{2}{N+1}} \sum_{k_z} \sin(k_z) \cc_{\bk \alpha \sigma},
\end{equation}
where the explicit form of the basis functions takes into account the
open boundary conditions which we impose on the system. Using a vector
representation for the orbitals fermionic operators:
$\hat{c}^{\dagger}_{\bk_\| z \sigma} \equiv \left(\cc_{\bk_\| z 1 \sigma},~ \cc_{\bk_\| z 2 \sigma} \right)$ 
and setting to unity the elementary charge $e=1$ we obtain
\begin{equation}
  \begin{split}
    \mathcal{H} &= 
    \sum_{\kpar \sigma} \sum_{z=1}^{N} 
    \hcc_{\bk_\| z \sigma} 
    \cdot 
    \bm{h}_{\kpar}
    \cdot   
    \hca_{\bk_\| z \sigma} 
    +
    \sum_{\kpar \sigma} \sum_{z=1}^{N-1} 
    \hcc_{\bk_\| z \sigma} 
    \cdot 
    \bm{t}_{\kpar}
    \cdot
    \hca_{\bk_\| z+1 \sigma} + H.c. \\ 
    &\phantom{=}-
    \frac{\Delta}{2} \sum_{z=1}^N \sum_{i \in z} 
    \left( n_{i,z,1} - n_{i,z,2} \right) +
    \frac{U}{2} \sum_{z=1}^N \sum_{i \in z} (n_{iz} - 2)^2 
    - \sum_{z} \sum_{i \in z} V(z) n_{i,z},
  \end{split}
  \label{eq:hamiltonian}
\end{equation}
where $\Delta$ is the crystal-field splitting, $U$ is the local Coulomb interaction strength, 
$V(z)$ is the scalar potential defined by Eq.~\ref{eq:scalar} and the matrices $\bm{h}_{\kpar}$ and
$\bm{t}_{\kpar}$ contain respectively the intra- and inter- layer hopping amplitudes
\begin{equation}
  \begin{split}
    &
    \bm{h}_{\kpar} =
    \left(
      \begin{matrix}
        \epsilon_{\kpar} & 0 \\
        0 & \epsilon_{\kpar}
      \end{matrix}
    \right),
    \quad
    \bm{t}_{\kpar} =
    \left(
      \begin{matrix}
        -t & v_{\kpar}  \\
        v_{\kpar} & -t
      \end{matrix}
    \right)\\
    &
    \text{with}
    \quad
    \left\lbrace
      \begin{matrix}
        \epsilon_{\kpar} =& -2t \left( \cos k_x + \cos k_y \right) \\
        v_{\kpar} =& v \left( \cos k_x - \cos k_y \right)
      \end{matrix}    
    \right.
    \end{split}
\end{equation}
The model is solved using the extension of the DMFT formalism to
in-homogeneous systems \cite{potthoff_rDMFT}, based 
on the assumption that the self-energies encoding the effect of all the many-body correlation 
are completely local in space while retaining an explicit dependence on the layer index $z$
\begin{equation}  
  \bm{\Sigma}_{iz,jz'}(i\omega_n) = \delta_{ij} \delta_{z z'}\bm{\Sigma}_z (i\omega_n).
\end{equation}
Following the standard DMFT approach, the layer-dependent self-energies are then
extracted from a set of layer-dependent single-site effective problems  
to be self-consistently determined imposing that the lattice local Green's function  $\bm{{G}}_{zz}^{-1}(i\omega_n)$
computed using such local self-energy is equal to the Green's function of the effective
single-site problem.
Indicating with $\bm{\mathcal{G}}^{-1}_{0,z}(i\omega_n)$ the bare propagators of the 
effective single-sites problems, the problem is practically solved using the following equations which implicitly relate
$\bm{{G}}_{zz}^{-1}(i\omega_n)$ and $\bm{\mathcal{G}}^{-1}_{0,z}(i\omega_n)$
\begin{equation}
  \begin{split}
    & \bm{G}_{zz} (i\omega_n) = \sum_{\kpar} \bm{G}_{\kpar zz}
    (i\omega_n), \\
    & \bm{{G}}_{zz}^{-1}(i\omega_n) =
    {\bm{\mathcal{G}}}^{-1}_{0,z}(i\omega_n) -
    {\bm{\Sigma}}_{z}(i\omega_n)
  \end{split}
\end{equation}
\begin{equation}
  \begin{split}
  \left[\bm{\hat{G}}(i\omega_n)\right]^{-1}_{zz'}\! 
  & =
  \delta_{zz'}
  \left[
    i \omega_n\mathbb{I}- \bm{h}_{\kpar} -  \bm{\Sigma}_z(i \omega_n) 
  \right] -
  \delta_{z,z \pm 1} \bm{t}_{\kpar},
  \end{split}
\end{equation}
where we indicate with $\bm{\hat{G}}(i \omega_n)$ the $2N \times 2N$ matrix constructed with all the $2 \times 2$ 
$\bm{G}_{zz'}(i \omega_n)$ matrices.
In the present case we map the effective single-sites problems onto interacting Anderson impurity models
which we solve using a finite bath discretization and an Exact Diagonalization 
scheme based on the Lanczos method~\cite{Review_DMFT_96}.

\begin{figure}
 \centering
 \hspace{0.5cm}
 \includegraphics[width=5.5cm]{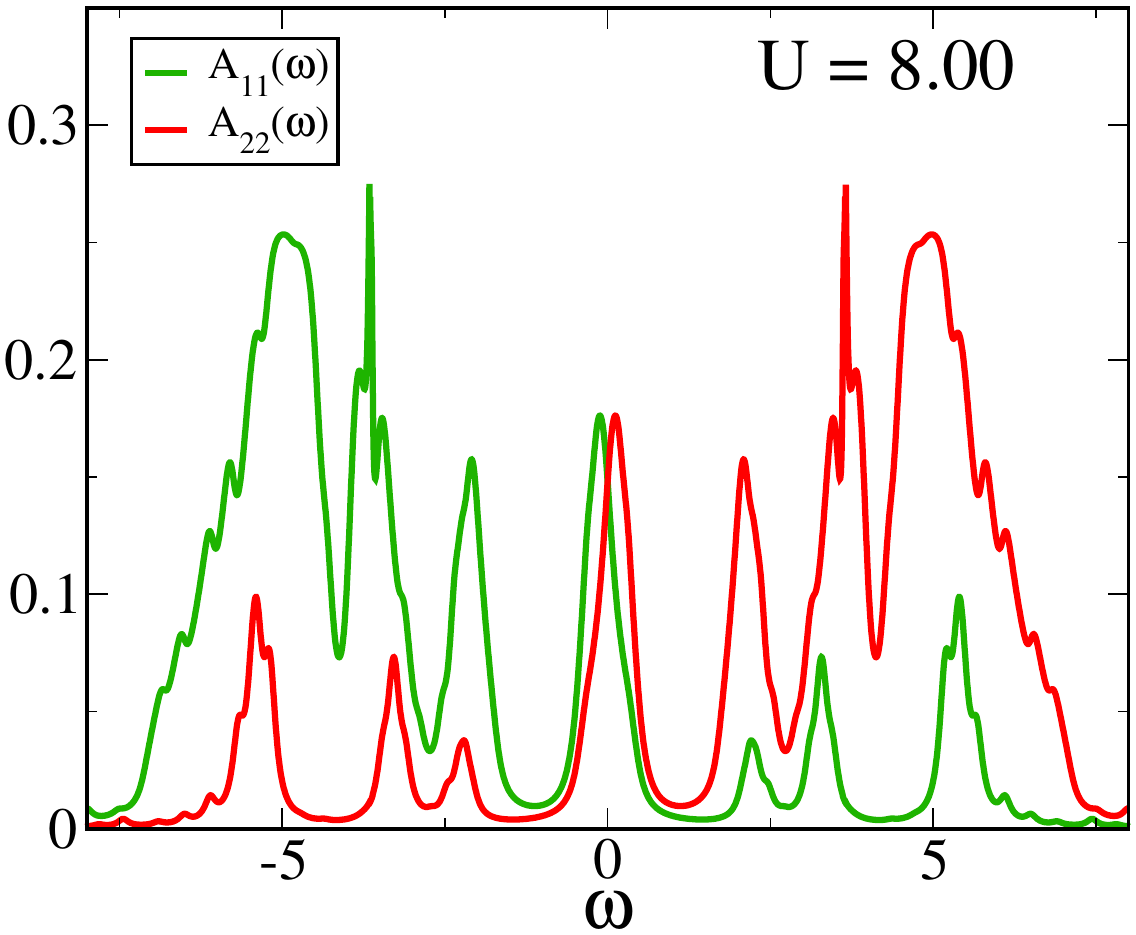}
 \hspace{1.5cm}
 \includegraphics[width=5.5cm]{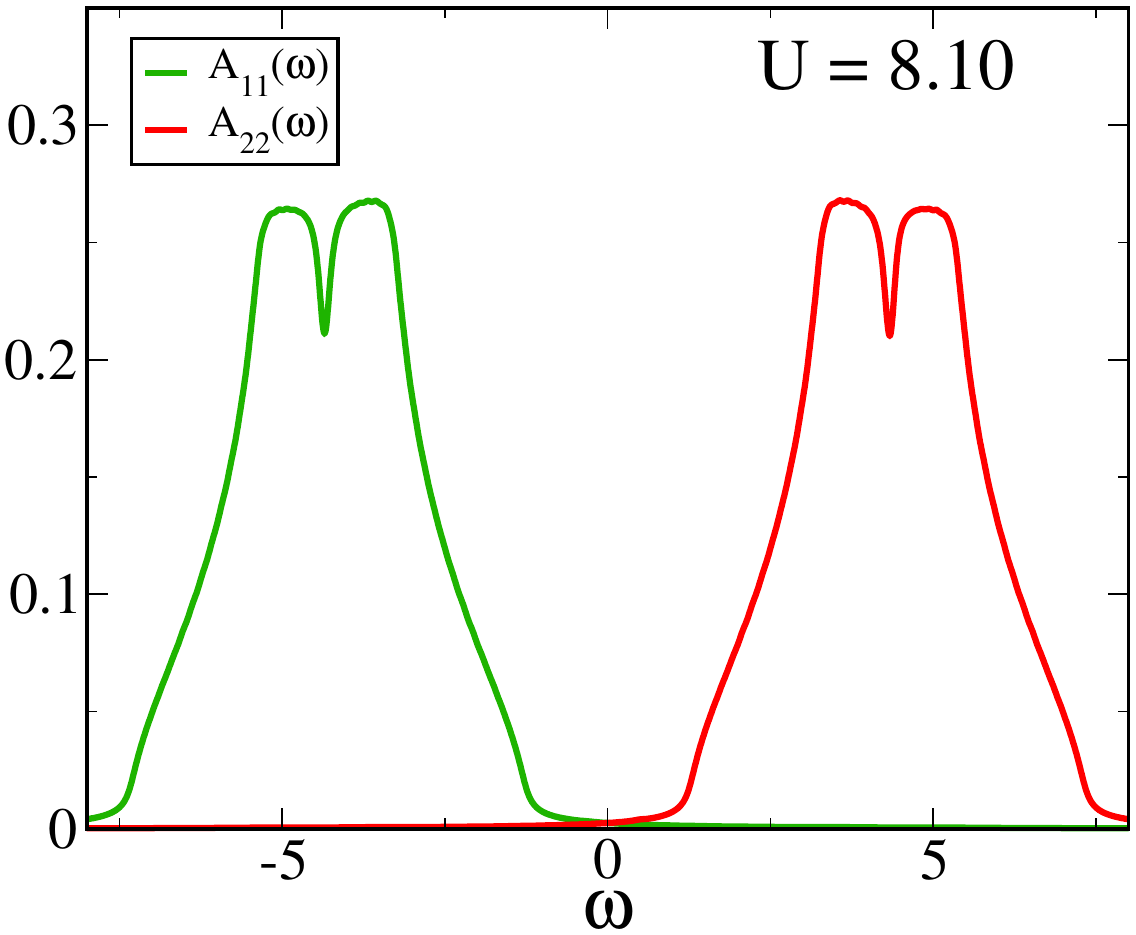}
 \caption{Orbital resolved spectral functions for the model Hamiltonian \ref{MainModel} across the first-order Mott
   transition. On the metallic side (left panel) the two spectral functions overlaps with a 
   partial orbital polarization ($m<2$). 
   On the insulating side (right panel) the two spectral functions are separated
   with an almost completed orbital polarization ($m \lesssim 2$).}
 \label{Fig1S} 
\end{figure}

\section{Zero-bias Mott transition}


In this section we provide few additional information concerning the zero bias metal-insulator transition.
As described in the main text the transition is driven by the combined
effects of the interaction enhancement  of the polarization strenght
and the shrinking of the coherent quasiparticle peak.  As a
consequence, the Mott transition appears as a sharp charge
redistribution between the two orbitals.  
We show this plotting in Fig.~\ref{Fig1S}(a) the orbital-resolved spectral functions 
$A_{\alpha \alpha}(\omega) = -\frac{1}{\pi} G_{\text{loc}}^{\alpha
  \alpha} (\omega)$ for two values of the interaction parameter just
before and after the metal-insulator  critical value $U_c \approx 8.05$. 
For $U \lesssim U_c$ the two spectral functions overlap with a sizable
spectral weight at the Fermi level displaying a two-orbitals
character. The abrupt separation of the orbital resolved spectral
weight occours for $U \gtrsim U_c$ leading to a Mott insulator in
which the lower band is fully occupied and the upper one empty (panel b Fig.~\ref{Fig1S}). 
We notice that due to the finite hybridization in the kinetic  part of
the Hamiltonian the orbital polarization is not exaclty complete with
less than one percent residual occupation  in the upper
orbital~\cite{poterayev_2bands}.

\section{Resistive- Vs. tunnel-like formation of the metallic states}
A more exhaustive description of the sharp differences between the
resistive- and the tunnel-breakdown described in  the main text can be
appreciated looking at the layer distributions of the charge density
and of the bias-induced spectral  density at the Fermi (see Fig.~\ref{Fig1S}(b)).

\begin{figure}
 \centering
  \includegraphics[width=5.5cm]{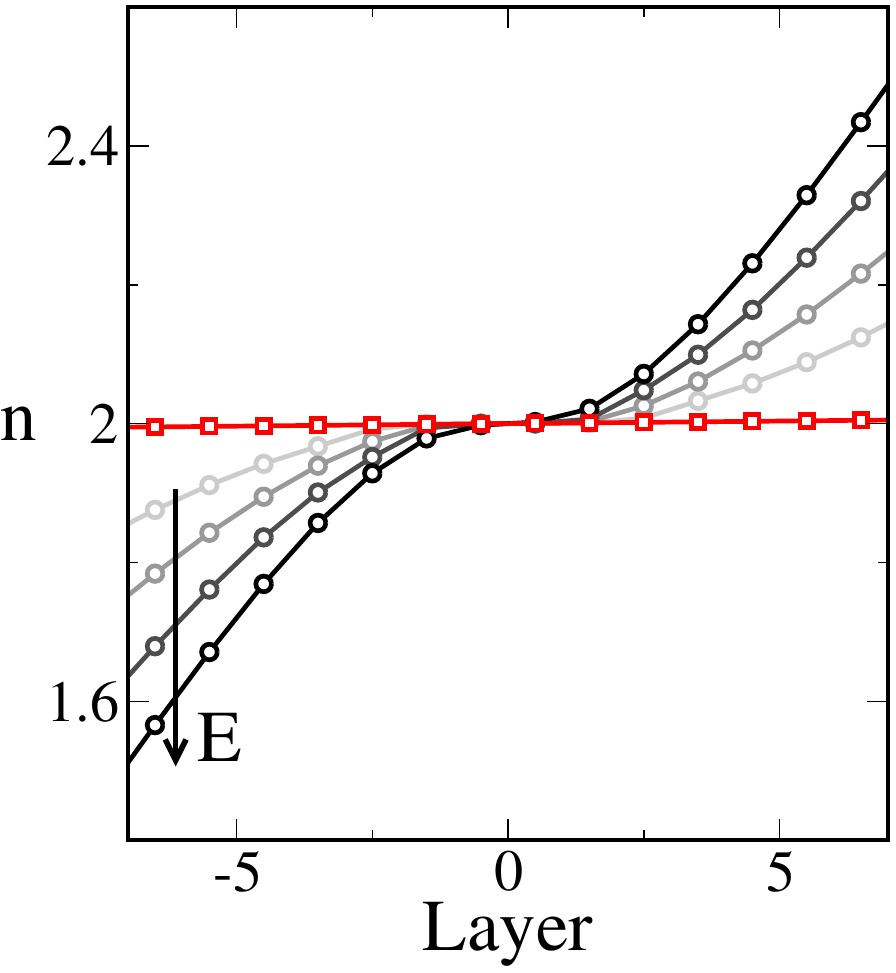}
  \hspace{1.5cm}
  \includegraphics[width=5.5cm]{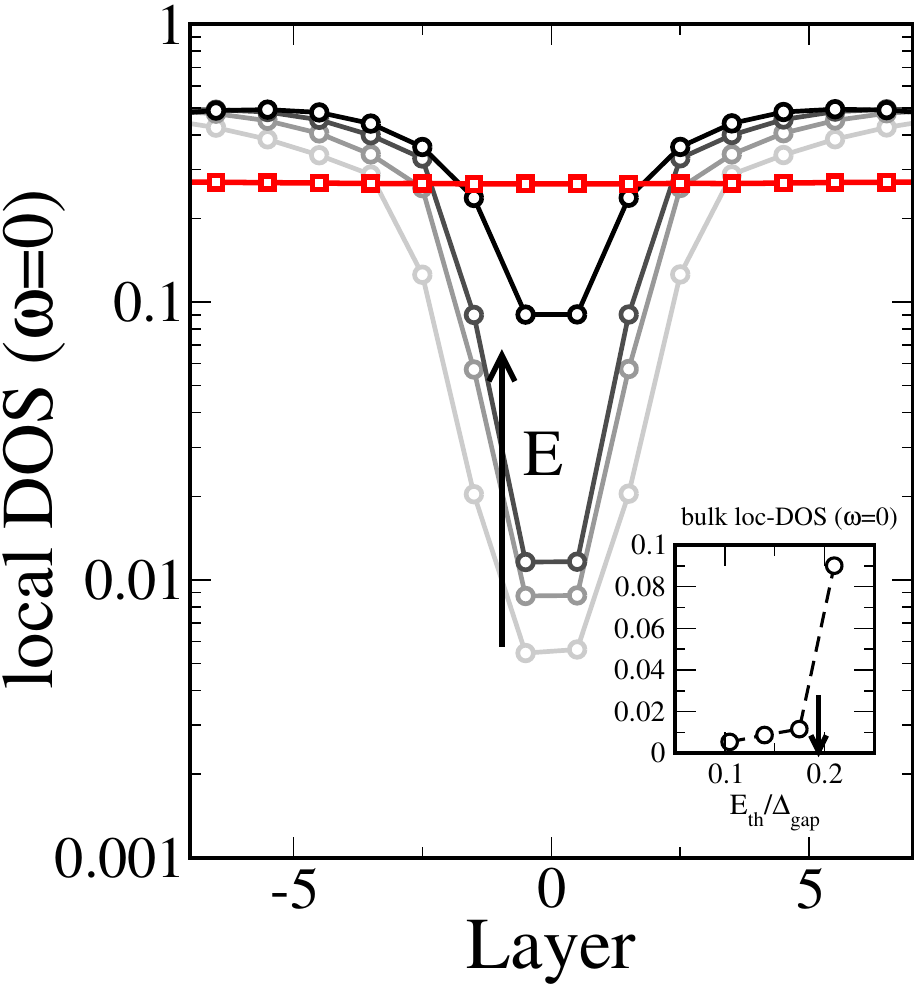}
   \caption{ 
     Layer density profile (left) and layer distribution of the
     spectral density at the Fermi level (right) for the bias-induced
     metallic states both for the resistive- (red squares) and
     tunneling- (gray to black circles) breakdown. For the resistive
     transition we show quantities (red squares) just after the
     transition while for the tunnel-breakdown (gray to black circles)
     we show the evolution as a function of the applied field (see
     arrows). The inset shows the evolution of the spectral densities
     at the fermi level for the bulk layers as a function of the
     applied field. From the finite jump we estimate the breakdown
     threshold field reported in the main text (small arrow).
 }
 \label{Fig2S}
 \end{figure}

In the case of the resistive switch, namely the insulator-to-metal
transition followed by the abrupt Mott gap collapse  close to the Mott
transition, we observe an infinitesimal tilting of the charge
distribution (red squares in Fig.~\ref{Fig2S}(b) left) and an
homogenous distribution of the spectral weight (red squares in
Fig.~\ref{Fig2S}(b) right).

On the contrary far away from the Mott transition, as expected by the
strong tilting of the layer-resolved spectral  density (main text), a
very strong charge redistribution is needed in order to observe the
formation of conducting states (gray to black circles in
Fig.~\ref{Fig2S}(b) left). 
This leads to the gradual formation of two metallic regions at the
boundaries which are separated by a bulk region in which the spectral
weight results exponentially suppressed (gray to black circles in
Fig.~\ref{Fig2S}(b) right). Such evanescent spectral weight represents
the tunnel through the bulk Mott insulating region of the
carriers from the two doped metallic regions at the slab boundaries~\cite{borghi_prb10}. 
This confirms the tunnel-like scenario for the field-induced formation 
of conducting paths deep in the Mott insulating phase (Fig.1e in the main text). 
In the inset we show the field evolution of the spectral density at the Fermi 
level for the bulk layer. The spectral density increases linearly up to a threshold
value where a clear jump is observed signaling the disappearence of
the bulk insulating region. We take this value to extract the breakdown threshold field 
for $U>U_s$ reported in the main text.

\end{document}